\documentclass[aps,pre,twocolumn,groupedaddress,showpacs]{revtex4-2}
\usepackage{graphicx}
\usepackage{color}
\usepackage{amsmath}
\usepackage{bm}
\usepackage{txfonts}

\begin{document}

\title{
Hierarchical Onsager symmetries in adiabatically driven linear irreversible heat engines
}

\author{Yuki Izumida}
\affiliation{Department of Complexity Science and Engineering, Graduate School of Frontier Sciences, The University of Tokyo, Kashiwa 277-8561, Japan}
\thanks{izumida@k.u-tokyo.ac.jp}
%%\homepage[]{Your web page}
%%\thanks{}
%%\altaffiliation{}

\begin{abstract}
In existing linear response theories for adiabatically driven cyclic heat engines, Onsager symmetry is identified only phenomenologically, and a relation between global and local Onsager coefficients, defined over one cycle and at any instant of a cycle, respectively, is not derived.
To address this limitation, we develop a linear response theory for the speed of adiabatically changing parameters and temperature differences in generic Gaussian heat engines obeying Fokker--Planck dynamics.
We establish a hierarchical relationship between the global linear response relations, defined over one cycle of the heat engines, and the local ones, defined at any instant of the cycle.
This yields a detailed expression for the global Onsager coefficients in terms of the local Onsager coefficients.
Moreover, we derive an efficiency bound, which is tighter than the Carnot bound, for adiabatically driven linear irreversible heat engines based on the detailed global Onsager coefficients.
Finally, we demonstrate the application of the theory using the simplest stochastic Brownian heat engine model.
\end{abstract}

\maketitle

{\sl Introduction}--.
The Carnot efficiency is the fundamental bound for the efficiency of heat engines, and it is universally imposed by equilibrium thermodynamics~\cite{Callen1985}.
Particularly, the Carnot efficiency is attained in an idealized reversible limit; however, the operation of actual powerful heat engines is accompanied by irreversible flows, and thus, should obey the constraints entailed by nonequilibrium thermodynamics.
The recent developments in understanding the constraints on nonequilibrium heat engines, including the finite-time thermodynamics~\cite{Berry2000,Salamon2001,Andersen2011}, 
the universality of efficiency at maximum power~\cite{Curzon1975,VandenBroeck2005,Borja2007,Schmiedl2008,Esposito2009,Esposito2010,Benenti2011,Izumida2014,Cavina2017}, the trade-off relation between power and efficiency~\cite{Brandner2013,Campisi2016,Raz2016,Shiraishi2016,Polettini2017,Pietzonka2018,Dechant2018,Holubec2018,Abiuso2020_2}, and geometrical formulations~\cite{Brandner2020,Abiuso2020_1,Miller_arxiv,Miller2020,Hino2021}, 
uncovered the universal features governing nonequilibrium heat engines beyond the Carnot efficiency.

Linear irreversible thermodynamics is a universal framework that systematically describes the response of equilibrium systems under weak nonequilibrium perturbations~\cite{Onsager1931,Callen1957}.
Despite its importance, the application of linear irreversible thermodynamics to heat engines operating under small temperature differences has been limited, until recently~\cite{Izumida2009,Izumida2010,Izumida2015,Brandner2015,Proesmans2015,Brandner2016,Proesmans2016_1,Proesmans2016_2,Cerino2016}.
This is because the identification of thermodynamic fluxes and forces is highly complex for heat engines undergoing cyclic changes.
Nevertheless, such an identification is essential because the performance of heat engines depends on the response coefficients, that is, Onsager coefficients, in the linear response regime~\cite{VandenBroeck2005,Benenti2011}.
In particular, the linear irreversible thermodynamics for the temperature difference and the speed of adiabatically changing parameters~\cite{adiabatic_change} of cyclic heat engines is limited to a few specific examples~\cite{Izumida2009,Izumida2010,Izumida2015}.
Adiabatically driven cyclic heat engines can experience continuous equilibrium change along a cycle and be substantially perturbed from a reference equilibrium point. This makes the application of the linear response theory, which is usually defined for a response from a one-equilibrium point, difficult and obscure.
Notably, the identified Onsager symmetry for these models is derived only phenomenologically, by adopting intuitive {\it global} fluxes and forces per cycle, without deriving a relation to the {\it local} thermodynamic fluxes and forces defined at any instant of a cycle.

By contrast, in recent studies on quantum thermoelectrics, such a linear response for adiabatically changing parameters 
has been investigated as an effect of adiabatic {\it ac} driving applied to a system~\cite{Ludovico2016,Bhandari2020}. 
Remarkably, the Onsager coefficients defined globally for a one-cycle period of ac driving, which determine the overall performance of the thermoelectrics,
are expressed in terms of locally defined Onsager coefficients at any instant during driving~\cite{Ludovico2016,Bhandari2020}.
The key of this formulation is to apply the standard linear response theory to instantaneous equilibrium states specified by 
the adiabatically changing parameters that are regarded to have ``frozen," fixed values.
Considering the universal nature of linear irreversible thermodynamics, 
we are motivated to uncover a similar hierarchical structure for adiabatically driven linear irreversible heat engines. 
To this end, we focus on the simplest heat engine model. 
We establish a hierarchical relationship between global and local Onsager coefficients for a generic Gaussian heat engine model obeying Fokker--Planck dynamics.
The adiabatic dynamics can be easily obtained based on the idea of time-scale separation~\cite{Strogatz2001}, which is one of the advantages of this model.
Moreover, based on the detailed structure of the Onsager coefficients, we derive an efficiency bound, tighter than the Carnot efficiency, under a given speed of adiabatic change.

{\sl Model}--.
The heat engine consists of a working substance (system) and thermal bath.
The state of the system $\mathbf x=(x_1, \cdots, x_n)$ at time $t$ is specified by a probability distribution $\mathcal{P}(\mathbf x,t)$.
The system is periodically operated based on $p$ external parameters ${\boldsymbol \lambda}(t)=(\lambda_1(t), \cdots, \lambda_p(t))$ and the bath temperature $T(t)$ with period $\tau_{\rm cyc}$; $\boldsymbol \lambda(t+\tau_{\rm cyc})=\boldsymbol \lambda(t)$ and $T(t+\tau_{\rm cyc})=T(t)$.
The energy of the system is given by $H(\mathbf x, t)$, which is a function of $\boldsymbol \lambda(t)$. 
Specifically, the external parameters are expressed as $\boldsymbol \lambda(t)=\boldsymbol \lambda_0+\mathbf g_w(\epsilon t)$ 
using the time-independent part $\boldsymbol \lambda_0$ and the time-dependent part $\mathbf g_w$. 
Here, $\epsilon \equiv 1/\tau_{\rm cyc}$ denotes a small parameter corresponding to the speed of the process.
Thus, a long period of time $t=O(1/\epsilon)$ is required for a finite increment of $\mathbf g_w$.
The bath temperature $T(t)$ is given by $T(t)=\frac{T_hT_c}{T_h-\Delta T(t)}$, where $\Delta T(t)\equiv \gamma_q(\epsilon t)\Delta T$, and
$\Delta T\equiv T_h-T_c$ and $\gamma_q(\epsilon t)$ are the temperature difference and periodic function satisfying $0\le \gamma_q(\epsilon t) \le 1$, respectively~\cite{Brandner2015}.

We define the average entropy production rate per cycle $\dot \sigma$ for the system and thermal bath. 
Hereafter, we denote by the overdot a quantity per unit time or a quantity being time differentiated.
The energy change rate becomes $\dot E\equiv \frac{d}{dt}\left<H(\mathbf x,t)\right>=\frac{d}{dt}\int d\mathbf x^n H(\mathbf x,t)\mathcal{P}(\mathbf x,t)$, where 
$\left<\cdot \right>$ refers to an ensemble average with respect to $\mathcal{P}(\mathbf x, t)$. 
We decompose $\dot E$ into the sum of the heat and work fluxes $\dot Q$ and $\dot W$; 
$\dot E=\int d\mathbf x^n H(\mathbf x,t)\frac{\partial \mathcal{P}(\mathbf x,t)}{\partial t}+\int d\mathbf x^n \frac{\partial H(\mathbf x,t)}{\partial t}\mathcal{P}(\mathbf x,t)\equiv \dot Q-\dot W$.
Then, we can define $\dot \sigma$ as
\begin{eqnarray}
\dot \sigma&&\equiv -\frac{1}{\tau_{\rm cyc}}\int_0^{\tau_{\rm cyc}} \frac{\dot Q(t)}{T(t)}dt\nonumber\\
&&=\frac{\epsilon}{T_c} \frac{1}{\tau_{\rm cyc}}\int_0^{\tau_{\rm cyc}} dt \int d^n \mathbf x \mathbf g'_w(\epsilon t)\cdot \frac{\partial H(\mathbf x, t)}{\partial \boldsymbol \lambda}\mathcal{P}(\mathbf x,t)\nonumber\\
&&+\left(\frac{1}{T_c}-\frac{1}{T_h}\right)\frac{1}{\tau_{\rm cyc}}\int_0^{\tau_{\rm cyc}} dt \int d^n \mathbf x \gamma_q(\epsilon t) H(\mathbf x, t) \dot {\mathcal{P}}(\mathbf x,t)\nonumber\\
&&=J_wF_w+J_qF_q,\label{eq.entropy_product}
\end{eqnarray}
where the prime symbol denotes the time derivative with respect to the {\it slow time} $\mathcal{T}\equiv \epsilon t$ and $\dot {\mathbf g}_w(\epsilon t)=\frac{d\mathbf g_w(\epsilon t)}{dt}=\epsilon \mathbf g'_w(\epsilon t)$. The dot between symbols denotes an inner product.
Here, we have defined the following work and heat fluxes per cycle as thermodynamic fluxes:
\begin{eqnarray}
J_w&&\equiv \frac{1}{\tau_{\rm cyc}}\int_0^{\tau_{\rm cyc}} dt \int d^n \mathbf x \mathbf g'_w(\epsilon t)\cdot \frac{\partial H(\mathbf x, t)}{\partial \boldsymbol \lambda} \mathcal{P}(\mathbf x,t),\label{eq.jw_def}\\
J_q&&\equiv \frac{1}{\tau_{\rm cyc}}\int_0^{\tau_{\rm cyc}} dt \int d^n \mathbf x \gamma_q(\epsilon t) H(\mathbf x, t) \dot{\mathcal{P}}(\mathbf x,t).\label{eq.jq_def}
\end{eqnarray}
The corresponding thermodynamic forces are defined as
\begin{eqnarray}
F_w\equiv \epsilon/T_c, \ \ F_q\equiv 1/T_c-1/T_h.
\end{eqnarray}
We assume the {\it global} linear response relations $\mathbf J=\mathrm L \mathbf F$ between $\mathbf J\equiv (J_w, J_q)^{\rm T}$ and $\mathbf F\equiv (F_w,F_q)^{\rm T}$ defined over one cycle of the heat engine in the limit of $\epsilon \to 0$ and $\Delta T\to 0$:
\begin{eqnarray}
&&J_w=L_{ww}F_w+L_{wq}F_q,\label{eq.jw}\\
&&J_q=L_{qw}F_w+L_{qq}F_q,\label{eq.jq}
\end{eqnarray}
where $\mathrm L$ corresponds to the global Onsager coefficients.
Our goal is to find a detailed expression of $\mathrm L$ in terms of its {\it local} counterpart defined at any instant of the cycle, thereby
establishing a hierarchical relationship between the two.

{\sl Fokker--Planck dynamics}--.
For further calculation of $\mathbf J$, we need to specify the dynamics of $\mathcal{P}(\mathbf x,t)$.
In what follows, we consider generic Gaussian heat engines described based on multivariate Ornstein--Uhlenbeck processes as the simplest models.
The energy of the system, which serves as a potential function,  thus takes the following quadratic form:
\begin{eqnarray}
H(\mathbf x, t)=\frac{1}{2}\mathbf x^{\mathrm{T}}\mathrm H(t) \mathbf x=\frac{1}{2}{\mathrm H}_{ij}(t)x_ix_j,
\end{eqnarray}
where $\mathrm H(t)$ is a positive-definite symmetric matrix ($i, j=1, \cdots, n$).
We assume that $\mathbf x$ is even variables under time reversal.
The probability distribution of the system $\mathcal{P}(\bm x, t)$ obeys the Fokker--Planck (FP) equation with the time-dependent drift matrix $A$ and diffusion matrix $B$~\cite{Vankampen2007,Gardiner2008}:
\begin{eqnarray}
\frac{\partial \mathcal{P}(\mathbf x, t)}{\partial t}
&&=-\frac{\partial}{\partial x_i}\left[A_{ij}(t)x_j \mathcal{P}(\mathbf x, t)-\frac{1}{2}B_{ij}(t)\frac{\partial \mathcal{P}(\mathbf x, t)}{\partial x_j}\right]\nonumber\\
&&=-\frac{\partial \mathcal{J}_i(\mathbf x,t)}{\partial x_i},\label{eq.fp}
\end{eqnarray}
where $\mathcal{J}_i(\mathbf x,t)$ is a probability current. 
$A$ is a symmetric matrix and $B$ is a positive-definite symmetric matrix. $B$ is further assumed to be invertible.
The probability distribution is assumed to be the zero-mean Gaussian distribution:
\begin{eqnarray}
\mathcal{P}(\mathbf x,t)=\frac{1}{(2\pi)^{n/2}}\frac{1}{\sqrt{\det \Xi(t)}}e^{-\frac{1}{2}\mathbf x^{\mathrm T}\Xi^{-1}(t)\mathbf x},\label{eq.gaussian}
\end{eqnarray}
where the symmetric covariance matrix $\Xi_{ij}\equiv \left<x_i x_j\right>-\left<x_i\right>\left<x_j\right>=\left<x_i x_j\right>$ obeys~\cite{Vankampen2007}
\begin{eqnarray}
\partial_t \Xi=2A\Xi+B.\label{eq.sigma}
\end{eqnarray}
Note that we assume that $A$ and $\Xi$ are commutative for simplicity.
The equation to be solved is replaced with the dynamical equations in Eq.~(\ref{eq.sigma}), instead of the FP equation in Eq.~(\ref{eq.fp}): 
Note that $A$, $B$, and $\mathrm H$ are not independent. 
For the time-independent energy $H(\mathbf x,t)=H_0(\mathbf x)$ and temperature $T(t)=T_c$, we have $A(t)=A_0$ and $B(t)=B_0$.
Then, the stationary solution $\Xi_0$ obtained as the solution of $\partial_t \Xi_0=0$ in Eq.~(\ref{eq.sigma}) satisfies
\begin{eqnarray}
2A_0=-B_0 \Xi_0^{-1}.\label{eq.Xi_st}
\end{eqnarray}
For the stationary distribution to agree with a Boltzmann distribution at temperature $T_c$, 
the following detailed balance condition is usually imposed~\cite{Gardiner2008}:
\begin{eqnarray} 
2A_0=-B_0 \frac{\mathrm H_0}{k_{\rm B}T_c},\label{eq.DB1}
\end{eqnarray}
which together with Eq.~(\ref{eq.Xi_st}) yields $\Xi_0^{-1}=\mathrm H_0/k_{\rm B}T_c$ with $k_{\mathrm B}$ being Boltzmann constant.
Here, as a natural generalization of Eq.~(\ref{eq.DB1}), we impose the detailed balance condition, including the time-dependent part: 
\begin{eqnarray} 
2A(t)=-B(t)\frac{\mathrm H(t)}{k_{\rm B}T(t)},\label{eq.DB2}
\end{eqnarray}
whose validation will be clarified below.

We decompose $A(t)$, $B(t)$, and $\Xi(t)$ into time-independent and time-dependent parts as 
$A(t)=A_0+\delta A(t)$, $B(t)=B_0+\delta B(t)$, and $\Xi(t)=\Xi_0+\delta \Xi(t)$.
Then, Eq.~(\ref{eq.sigma}) is replaced with
\begin{eqnarray}
\partial_t \delta \Xi=2A(t)\delta \Xi+2\delta A(t)\Xi_0+\delta B(t).\label{eq.deltasigma}
\end{eqnarray}
We solve Eq.~(\ref{eq.deltasigma}) perturbatively with respect to $\epsilon$.
Because a regular perturbation yields a secular term, we use a two-timing method based on time-scale separation~\cite{Strogatz2001}.
As a result, we obtain $\Xi(t)$ as (see Supplemental Material~\cite{SM})
\begin{eqnarray}
\Xi(t)=\Xi_0+\delta \Xi(t)=\Xi_{\rm ad}(t)+\delta \Xi_{\rm nad}(t)+O(\epsilon^2),\label{eq.xi}
\end{eqnarray}
where $\Xi_{\rm ad}(t)$ and $\delta \Xi_{\rm nad}(t)$ are the adiabatic solution and the lowest non-adiabatic correction to it, respectively, as
\begin{eqnarray}
&&\Xi_{\rm ad}(t)\equiv -\frac{1}{2}A^{-1}(t)B(t),\label{eq.xi_ad}\\
&&\delta \Xi_{\rm nad}(t)\equiv -\Xi_{\rm ad}B^{-1}(t)\frac{\partial \Xi_{\rm ad}}{\partial \mathcal{T}}\epsilon. \label{eq.xi_nad}
\end{eqnarray}
From Eqs.~(\ref{eq.DB2}) and (\ref{eq.xi_ad}), we have
\begin{eqnarray}
\Xi_{\rm ad}^{-1}(t)=\frac{\mathrm H(t)}{k_{\mathrm B}T(t)}.\label{eq.H}
\end{eqnarray}
Thus, the probability distribution $\mathcal P(\mathbf x,t)$ in the adiabatic limit $\epsilon \to 0$ agrees with an instantaneous equilibrium distribution with energy $H(\mathbf x, t)$ and temperature $T(t)$, which validates the condition given by Eq.~(\ref{eq.DB2}).

{\sl Local and global linear response relations for speed and temperature differences}--.
We can now evaluate the thermodynamic fluxes in Eqs.~(\ref{eq.jw_def}) and (\ref{eq.jq_def}) using Eqs.~(\ref{eq.xi})--(\ref{eq.xi_nad}).
Note that we can rewrite Eq.~(\ref{eq.jq_def}) as $J_q=\frac{1}{\tau_{\rm cyc}}\int_0^{\tau_{\rm cyc}}dt \gamma_q(\epsilon t) \int d\mathbf x^n \frac{\partial H(\mathbf x, t)}{\partial x_i}\mathcal{J}_i(\mathbf x,t)$ using Eq.~(\ref{eq.fp}), and we can express Eq.~(\ref{eq.jw_def}) and (\ref{eq.jq_def}) as the time average of the {\it local} thermodynamic fluxes as
\begin{eqnarray}
J_w&&=\frac{1}{\tau_{\rm cyc}}\int_0^{\tau_{\rm cyc}} dt \mathbf g'_w(\epsilon t)\cdot \bm  j_w(t),\label{eq.jw_fp}\\
J_q&&=\frac{1}{\tau_{\rm cyc}}\int_0^{\tau_{\rm cyc}}dt \gamma_q(\epsilon t)j_q(t),\label{eq.jq_fp}
\end{eqnarray}
respectively, where we define the response vectors $\mathbf j=(\bm j_w, j_q)^{\rm T}\equiv \left(\left<\frac{\partial H(\mathbf x, t)}{\partial \boldsymbol \lambda}\right>,\int d\mathbf x^n \frac{\partial H(\mathbf x, t)}{\partial x_i}\mathcal{J}_i(\mathbf x,t)\right)^{\rm T}$ as the local thermodynamic fluxes.
We also introduce the conjugate local nonequilibrium perturbation vector $\mathbf f=(\bm f_w, f_q)^{\rm T}\equiv (\dot{\boldsymbol \lambda}, \Delta T(t)/T_c)^{\rm T}=(\epsilon \mathbf g'_w, \gamma_q \Delta T/T_c)^{\rm T}$. 
The perturbations are the speed of adiabatically changing parameters 
and temperature difference, and the responses are the generalized pressure and instantaneous heat flux.
The relationship between the perturbations and responses can be written as a {\it local} flux-force form~\cite{Ludovico2016,Bhandari2020}, namely $\mathbf j=\mathbf j_{\rm ad}+\mathbf \Lambda \mathbf f$ to the linear order of $\mathbf f$,
where $\mathbf j_{\rm ad}$ is an adiabatic response that remains in the limit of $\epsilon \to 0$ and $\Delta T\to 0$, and $\mathbf \Lambda$ is the {\it local} Onsager matrix given by
\begin{eqnarray}
&&\mathbf \Lambda=
\left(
\begin{array}{cc}
\boldsymbol \Lambda_{ww} & \boldsymbol \Lambda_{wq}\\
\boldsymbol \Lambda_{qw} & \Lambda_{qq}
\end{array}
\right).\label{eq.Lambda}
\end{eqnarray}
We can expand $\bm j_w$ and $j_q$ with respect to $\mathbf f$ as (see Supplemental Material~\cite{SM})
\begin{eqnarray}
\bm j_w&&\simeq -\frac{k_{\rm B}T(t)}{2}\mathrm \Xi_{\rm ad}^{-1}\cdot \frac{\partial \mathrm \Xi_{\rm ad}}{\partial \boldsymbol \lambda}+\frac{k_{\rm B}T_c}{2}\frac{\partial \mathrm \Xi_{\rm ad}}{\partial \boldsymbol \lambda}\mathrm \Xi_{\rm ad}^{-1} \cdot B_0^{-1}\frac{\partial \mathrm \Xi_{\rm ad}}{\partial \boldsymbol \lambda}\cdot \epsilon \mathbf g'_w,\label{eq.jw_local}\\
j_q&&\simeq \frac{k_{\rm B}T_c}{2}\mathrm \Xi_{\rm ad}^{-1}\cdot \frac{\partial \mathrm \Xi_{\rm ad}}{\partial \boldsymbol \lambda}\cdot \epsilon \mathbf g'_w\label{eq.jq_local},
\end{eqnarray}
to the linear order of $\mathbf f$.
We thus identify $\mathbf j_{\rm ad}$ and $\mathbf \Lambda$ as
\begin{eqnarray}
&&\mathbf j_{\rm ad}=
\left(
\begin{array}{cc}
-\frac{k_{\rm B}T_c}{2}\mathrm \Xi_{\rm ad}^{-1}\frac{\partial \mathrm \Xi_{\rm ad}}{\partial \boldsymbol \lambda} \\
0
\end{array}
\right), \label{eq.j_ad}\\
&&\mathbf \Lambda=
\left(
\begin{array}{cc}
\frac{k_{\rm B}T_c}{2}\frac{\partial \mathrm \Xi_{\rm ad}}{\partial \boldsymbol \lambda}\mathrm \Xi_{\rm ad}^{-1}\cdot  B_0^{-1} \frac{\partial \mathrm \Xi_{\rm ad}}{\partial \boldsymbol \lambda} & -\frac{k_{\rm B}T_c}{2}\mathrm \Xi_{\rm ad}^{-1}\cdot \frac{\partial \mathrm \Xi_{\rm ad}}{\partial \boldsymbol \lambda}\\
\frac{k_{\rm B}T_c}{2}\mathrm \Xi_{\rm ad}^{-1}\cdot \frac{\partial \mathrm \Xi_{\rm ad}}{\partial \boldsymbol \lambda} & 0
\end{array}
\right),\label{eq.L_l}
\end{eqnarray}
respectively. 
We can confirm the Onsager symmetry $\Lambda_{ww, mm'}=\Lambda_{ww, m'm}$ and anti-symmetry $\Lambda_{wq, m}=-\Lambda_{qw, m}$ ($m, m'=1,\cdots, p$) at the local level.
The former symmetry relates to the dissipation, while the latter anti-symmetry relates to the dissipationless cross-coupling between the heat flux and the work flux (heat engine--refrigerator symmetry).

Subsequently, we consider the global linear response relations $\mathbf J=\mathrm L \mathbf F$ in Eqs.~(\ref{eq.jw}) and (\ref{eq.jq}).
The global thermodynamic fluxes in Eqs.~(\ref{eq.jw_fp}) and (\ref{eq.jq_fp}) can be rewritten as $J_w=\int_0^1d\mathcal{T} \mathbf g'_w(\mathcal{T})\cdot \bm j_w$ and $J_q=\int_0^1d\mathcal{T} \gamma_q(\mathcal{T})j_q$ in terms of the slow time $\mathcal T=\epsilon t$.
We note that the contribution from $\mathbf j_{\rm ad}$ vanishes upon cycle averaging.
Note that $F_w=\epsilon/T_c$ and $F_q\simeq \Delta T/T_c^2$ in the linear response regime, and using Eqs.~(\ref{eq.jw_local}) and (\ref{eq.jq_local}), we immediately arrive at the following expression for the global Onsager matrix $\mathrm L$:
\begin{eqnarray}
\mathrm L=
\left(
\begin{array}{cc}
T_c \int_0^1 d\mathcal{T}  \mathbf g'_w \cdot \mathbf \Lambda_{ww} \cdot \mathbf g'_w & T_c \int_0^1 d\mathcal{T} \gamma_q \mathbf \Lambda_{wq}\cdot \mathbf g'_w\\
T_c \int_0^1 d\mathcal{T} \gamma_q \mathbf \Lambda_{qw} \cdot \mathbf g'_w & 0
\end{array}
\right).\label{eq.L_g}
\end{eqnarray}
The local and global Onsager matrices in Eqs.~(\ref{eq.L_l}) and (\ref{eq.L_g}) constitute the first main results of this study.
The global Onsager coefficients $\mathrm L$ are given as the integration over one cycle of the local Onsager coefficients $\mathbf \Lambda$ in Eq.~(\ref{eq.L_l}).
This yields a hierarchical relationship between $\mathrm L$ and $\mathbf \Lambda$, thereby relating the different levels of symmetries.
In particular, $\mathrm L$ shows Onsager anti-symmetry $L_{wq}=-L_{qw}$, reflecting the Onsager anti-symmetry $\Lambda_{wq, m}=-\Lambda_{qw, m}$ for $\mathbf \Lambda$.

In the linear response regime, the average entropy production rate per cycle $\dot \sigma=J_wF_w+J_qF_q$ in Eq.~(\ref{eq.entropy_product}) takes the quadratic form 
$\dot \sigma=L_{ww}F_w^2+(L_{wq}+L_{qw})F_wF_q+L_{qq}F_q^2$,
 where we have used Eqs.~(\ref{eq.jw}) and (\ref{eq.jq}).
The second law of thermodynamics $\dot \sigma \ge 0$ imposes constraints on $\mathrm L$:
\begin{eqnarray}
L_{ww}\ge 0, \ L_{qq}\ge 0, \ L_{ww}L_{qq}-(L_{wq}+L_{qw})^2/4\ge 0.
\end{eqnarray}
For the present system, we find 
\begin{eqnarray}
\dot \sigma=L_{ww}F_w^2,
\end{eqnarray}
by using the explicit form of $\mathrm L$ in Eq.~(\ref{eq.L_g}).
Remarkably, we readily observe $L_{ww}\ge 0$, and thus, $\dot \sigma \ge 0$ from the positive-definite quadratic form of $L_{ww}$ in Eq.~(\ref{eq.L_g}).
The anti-symmetric coefficients do not contribute to $\dot \sigma$ because they represent a reversible, adiabatic change in entropy.
The vanishing $L_{qq}$ also reduces $\dot \sigma$, which arises from nonsimultaneous contact with the thermal baths at different temperatures.
This property is essentially the same as that known as the tight-coupling condition~\cite{VandenBroeck2005}. 
Note that we have the optional thermodynamic fluxes and forces. 
By switching the roles of $J_w$ and $F_w$, that is $\tilde{J}_w=F_w$ and $\tilde{F}_w=J_w$, while maintaining $\tilde{J}_q=J_q$ and $F_q=\tilde{F}_q$, we obtain another global Onsager matrix $\mathrm {\tilde L}$:
\begin{eqnarray}
\mathrm {\tilde L}=
\left(
\begin{array}{cc}
\frac{1}{L_{ww}} & -\frac{L_{wq}}{L_{ww}}\\
\frac{L_{qw}}{L_{ww}} & -\frac{L_{qw}L_{wq}}{L_{ww}}
\end{array}
\right),
\end{eqnarray}
assuming that $L_{ww}$ is nonvanishing and using $L_{qq}=0$.
Thus, we can confirm the symmetric non-diagonal elements and the vanishing determinant, where the latter corresponds to the tight-coupling condition.
Such a choice of fluxes and forces was adopted to identify Onsager coefficients of the finite-time Carnot cycle in~\cite{Izumida2009,Izumida2010,Izumida2015}.
As we will see below, the vanishing $L_{qq}$, equivalently, the tight-coupling condition, implies the attainability of the Carnot efficiency in the adiabatic limit $\epsilon \to 0$~\cite{Ludovico2016}.

{\sl Thermodynamic efficiency}--.
Using the global linear response relations in Eqs.~(\ref{eq.jw}) and (\ref{eq.jq}) together with Eq.~(\ref{eq.L_g}), we formulate the power $P$ and efficiency $\eta$ of our Gaussian heat engines:
\begin{eqnarray}
&&P\equiv -J_wF_wT_c=-(L_{ww}F_w+L_{wq}F_q)F_wT_c,\\
&&\eta \equiv \frac{P}{J_q}=\frac{-J_wF_wT_c}{J_q}=\eta_{\rm C}-\frac{L_{ww}}{L_{qw}}F_wT_c,
\end{eqnarray}
where $\eta_{\rm C}\equiv \Delta T/T_h \simeq \Delta T/T_c$ is the Carnot efficiency.
In the adiabatic limit $F_w \to 0$, we recover $\eta=\eta_{\rm C}$. 
For small $\epsilon$, the power behaves as $P=-L_{wq}\Delta T\epsilon/T_c^2+O(\epsilon^2)$. 
It should agree with $\Delta T\Delta S \epsilon$, where $\Delta S$ denotes an adiabatic entropy change of the system and $\Delta T \Delta S$ is an adiabatic work per cycle.
Thus, we identify $L_{wq}=-L_{qw}=-T_c^2 \Delta S$, which clarifies the vanishing contribution of these antisymmetric parts to the irreversible 
average entropy production rate $\dot \sigma$.
The efficiency under a given $F_w$, that is, the speed $\epsilon$, is bounded by the upper side as
\begin{eqnarray}
\eta \le \eta_{\rm C}-\frac{\mathcal{L}^2}{T_c\Delta S}\epsilon,\label{eq.eta_bound}
\end{eqnarray}
where $T_c \mathcal{L}^2$ is the minimum value of $L_{ww}$.
Reparameterizing from $\mathcal{T}$ to $\theta$ ($0\le \theta \le 1$), we have $\int_0^1 d\mathcal{T} \mathbf g'_w \cdot \mathbf \Lambda_{ww} \cdot  \mathbf g'_w=\int_0^1 d\mathcal{T} \frac{d\mathbf g_w}{d\theta} \cdot \mathbf \Lambda_{ww}\cdot \frac{d\mathbf g_w}{d\theta}|\theta'(\mathcal{T})|^2$.
Using the Cauchy--Schwartz inequality, we obtain 
$L_{ww} \ge T_c \left|\int_0^1 \sqrt{\frac{d\mathbf g_w}{d\theta} \cdot \boldsymbol \Lambda_{ww}\cdot \frac{d\mathbf g_w}{d\theta}} d\theta \right|^2\equiv T_c \mathcal{L}^2$~\cite{Sekimoto1997}.
Equation~(\ref{eq.eta_bound}) constitutes our second main result.
It yields a tighter bound than the Carnot efficiency imposed by the conventional second law of thermodynamics and is attained for an optimal protocol under a given cycle speed.
Such a bound was obtained by virtue of the detailed structure of the global Onsager coefficients (Eq.~(\ref{eq.L_g})).
$\mathcal{L}$ is equivalent to the thermodynamic length, 
which constrains the minimum dissipation along finite-time transformations close to equilibrium states~\cite{Salamon1983,Sekimoto1997,Crooks2007,Sivak2012,Zulkowski2012,Bonanca2014,Deffner2013,Deffner2020}.
An expression similar to Eq.~(\ref{eq.eta_bound}) including an effect of temperature-variation speed was recently derived based on a geometric formulation of quantum heat engines~\cite{Brandner2020}.
Here, we derived the similar form in terms of the global linear response relations between the speed of adiabatically changing parameters and temperature difference.

{\sl Example: Brownian heat engine}--.
We demonstrate our results by using the simplest illustrative case of a one-dimensional stochastic Brownian heat engine model ($n=p=1$)~\cite{Schmiedl2008,Raz2016,Sekimoto1997}.
Let $x_1=x$ be the position of a Brownian particle immersed in a thermal bath.
The probability $\mathcal{P}(x,t)$ obeys the following FP equation~\cite{Sekimoto2010,Seifert2012}:
\begin{eqnarray}
\frac{\partial \mathcal{P}(x,t)}{\partial t}=-\frac{\partial}{\partial x}\left[-\frac{1}{\gamma}\frac{\partial U(x,t)}{\partial x}\mathcal{P}(x,t)-\frac{k_{\mathrm B}T(t)}{\gamma}\frac{\partial \mathcal{P}(x,t)}{\partial x}\right],
\end{eqnarray}
where $\gamma$ is viscous friction coefficient and $H(x, t)=U(x,t)=\frac{\lambda(t)}{2}x^2$ with $\lambda(t)=\lambda_0+g_w(\epsilon t)$ is a harmonic potential.
We identify $A$ and $B$ as $A=A_{11}=-\frac{\lambda(t)}{\gamma}$ and $B=B_{11}=\frac{2k_{\mathrm B}T(t)}{\gamma}$.
Because the Boltzmann distribution with $T_c$ and $\lambda_0$ is $p_0(x)=\sqrt{\frac{\lambda_0}{2\pi k_{\rm B}T_c}}e^{-\frac{\lambda_0 x^2}{2k_{\mathrm B}T_c}}$, the variance at equilibrium is $\Xi_{0, 11}=k_{\mathrm B}T_c/\lambda_0$.

The adiabatic solution is given by $\Xi_{{\rm ad}, {11}}(t)=k_{\mathrm B}T(t)/\lambda(t)$.
The local linear response relations $\mathbf j=\mathbf j_{\rm ad}+\mathbf \Lambda \mathbf f$ are then obtained from Eqs.~(\ref{eq.j_ad}) and (\ref{eq.L_l}) as
\begin{eqnarray}
j_w=\frac{k_{\rm B}T(t)}{2\lambda(\epsilon t)}+\frac{\gamma k_{\rm B}T_c}{4\lambda^3(\epsilon t)}\epsilon g_w'(\epsilon t), \ j_q=-\frac{k_{\rm B}T_c}{2\lambda(\epsilon t)}\epsilon g'_w(\epsilon t), 
\end{eqnarray}
up to $O(\mathbf f)$, which determines the local and global Onsager matrices $\mathbf \Lambda$ and $\mathrm L$ as
\begin{eqnarray}
&&\boldsymbol \Lambda=
\left(
\begin{array}{cc}
\frac{\gamma k_{\rm B}T_c}{4\lambda^3(\epsilon t)} & \frac{k_{\rm B}T_c}{2\lambda(\epsilon t)} \\
-\frac{k_{\rm B}T_c}{2\lambda(\epsilon t)} & 0
\end{array}
\right),\\
&&\mathrm L=
\left(
\begin{array}{cc}
\gamma k_{\rm B}T_c^2 \int_0^1d\mathcal{T}\frac{{g'_w(\mathcal{T})}^2}{4\lambda^3(\mathcal{T})} & \frac{k_{\rm B}T_c^2}{2} \int_0^1 d\mathcal{T}\frac{g'_w(\mathcal{T})\gamma_q(\mathcal{T})}{\lambda(\mathcal{T})} \\
-\frac{k_{\rm B}T_c^2}{2} \int_0^1 d\mathcal{T}\frac{g'_w(\mathcal{T})\gamma_q(\mathcal{T})}{\lambda(\mathcal{T})} & 0
\end{array}
\right),
\end{eqnarray}
respectively. We can confirm the Onsager anti-symmetry in $\mathbf \Lambda$ and $\mathrm L$, as expected.
For a Carnot-like cycle with $\gamma_q(\mathcal{T})=1$ for $0\le \mathcal{T} < \mathcal{T}_h$ ($0<\mathcal T_h<1$) and $\gamma_q(\mathcal{T})=0$ for $\mathcal{T}_h \le \mathcal{T} \le 1$~\cite{Brandner2015}, we have 
$L_{wq}=-L_{qw}=-T_c^2\Delta S=\frac{k_{\rm B}T_c^2}{2}\ln (\lambda_1/\lambda_0)$, where $\lambda_1\equiv \lambda(\mathcal{T}_h)$ and $\lambda_0=\lambda(0)=\lambda(1)$ are the minimum and maximum values of $\lambda$ along the cycle, respectively.
We can obtain 
\begin{eqnarray}
\mathcal L^2=\frac{\gamma k_{\rm B}T_c}{\mathcal{T}_h(1-\mathcal{T}_h)}\left[\frac{1}{\sqrt{\lambda_1}}-\frac{1}{\sqrt{\lambda_0}}\right]^2
\end{eqnarray}
using the optimal protocol $\lambda^*(\mathcal{T})$ for a given $\lambda_0$ and $\lambda_1$~\cite{Sekimoto1997}: 
\begin{eqnarray}
\lambda^*(\mathcal{T})=
\begin{cases}
& \left[\frac{\mathcal{T}}{\mathcal{T}_h\sqrt{\lambda_1}}+\frac{\mathcal{T}_h-\mathcal{T}}{\mathcal{T}_h\sqrt{\lambda_0}}\right]^{-2} \ (0\le \mathcal{T} < \mathcal{T}_h),\\ 
%$\lambda^*(\mathcal{T})
&\left[\frac{\mathcal{T}-\mathcal{T}_h}{(1-\mathcal{T}_h)\sqrt{\lambda_0}}+\frac{1-\mathcal{T}}{(1-\mathcal{T}_h)\sqrt{\lambda_1}}\right]^{-2} 
\ (\mathcal{T}_h \le \mathcal{T} \le 1).
\end{cases}
\end{eqnarray}
The efficiency bound in Eq.~(\ref{eq.eta_bound}) for the present case thus becomes
\begin{eqnarray}
\eta_{\rm C}-\frac{2\gamma \left|\frac{1}{\sqrt{\lambda_1}}-\frac{1}{\sqrt{\lambda_0}}\right|^2}{\mathcal{T}_h(1-\mathcal{T}_h)\ln \left(\frac{\lambda_0}{\lambda_1}\right)}\epsilon.\label{eq.bound_bp}
\end{eqnarray}
A comparison of the bound given by Eq.~(\ref{eq.bound_bp}) with that, for example, using $L_{ww}=\frac{\gamma k_{\mathrm B}T_c^2(\lambda_0-\lambda_1)}{8\mathcal{T}_h(1-\mathcal{T}_h)}\left(\frac{1}{\lambda_1^2}-\frac{1}{\lambda_0^2}\right)$ for a linear protocol connecting $\lambda_0$ and $\lambda_1$ 
highlights the importance of protocol optimization as a design principle.

{\sl Concluding perspective}--.
We developed a linear response theory for generic Gaussian heat engines as the simplest model of adiabatically driven linear irreversible heat engines.
We established the hierarchical relationship between the local and global Onsager coefficients.
Further, we derived the efficiency bound under a given rate of adiabatic change; the derived bound is tighter than the Carnot efficiency imposed by the second law of thermodynamics.
We expect that the present results will contribute to a deeper understanding of the physical principles and optimal control of nonequilibrium heat engines.

We note complementary approaches for the formulation of the linear irreversible thermodynamics to periodically driven heat engines in ~Refs.~\cite{Brandner2015,Proesmans2015,Brandner2016,Proesmans2016_1,Proesmans2016_2,Cerino2016}.
In these approaches, the other thermodynamic force (that is, in addition to the temperature difference) is the strength of periodic forcing, and not its speed, as in the present approach. 
Interestingly, the Onsager coefficients in these cases were found to be decomposed into adiabatic and non-adiabatic contributions.
The existence of different types of linear irreversible thermodynamics implies the rich and versatile structures of periodically driven heat engines, and this deserves further investigation.

\begin{acknowledgements}
The author is grateful for valuable discussions at a seminar organized by K. Takahashi.
This work was supported by JSPS KAKENHI (Grant No. 19K03651).
\end{acknowledgements}

\clearpage
\onecolumngrid

\begin{center}
\textbf{\large Supplemental Material for ``Hierarchical Onsager symmetries in adiabatically driven linear irreversible heat engines"}
\end{center}

\begin{center}
Yuki Izumida\\
{\it Department of Complexity Science and Engineering, Graduate School of Frontier Sciences, The University of Tokyo, Kashiwa 277-8561, Japan}
\end{center}
\vspace{5pt}

\section{Derivation of Eqs.~(15)--(17)}
We solve Eq.~(14) in the main text perturbatively with respect to $\epsilon$ by using the two-timing method~\cite{Strogatz2001_SM}.
By introducing the slow time scale $\mathcal{T}\equiv \epsilon t$ and fast time scale $\tau \equiv t$, we expand $\delta \Xi$ as $\delta \Xi(t, \epsilon)=\delta \Xi^{(0)}(\tau, \mathcal{T})+\epsilon \delta \Xi^{(1)}(\tau, \mathcal{T})+O(\epsilon^2)$ by considering $\mathcal{T}$ and $\tau$ as independent variables. 
The differential operator is thus written as
$\frac{\partial \delta \Xi}{\partial t}=\frac{\partial \delta \Xi}{\partial \tau}+\epsilon \frac{\partial \delta \Xi}{\partial \mathcal{T}}$. 
By substituting $\delta \Xi(t, \epsilon)$ and $\partial_t \delta \Xi(t, \epsilon)$ into Eq.~(14), we obtain the following equation for each order of $\epsilon$:
\begin{align}
O(1): \frac{\partial \delta \mathrm \Xi^{(0)}}{\partial \tau}=2A(\mathcal{T})\delta \Xi^{(0)}+\delta F(\mathcal{T}),\tag{S1}\\
O(\epsilon): \frac{\partial \delta \mathrm \Xi^{(1)}}{\partial \tau}=2A(\mathcal{T})\delta \mathrm \Xi^{(1)}-\frac{\partial \delta \mathrm \Xi^{(0)}}{\partial \mathcal{T}},\tag{S2}
\end{align}
where $\delta F(t)\equiv 2\delta A(t)\Xi_0+\delta B(t)$.
We first solve Eq.~(S1). By solving the homogeneous differential equation in Eq.~(S1), we obtain
$\delta \Xi^{(0)}=e^{2 A(\mathcal{T})\tau}R(\tau)$, 
where $R(\tau)$ satisfies $\frac{\partial R(\tau)}{\partial \tau}=e^{-2A(\mathcal{T})\tau}\delta F(\mathcal{T})$, using $\left(e^C\right)^{-1}=e^{-C}$ for an invertible matrix $C$, 
which can be solved as $R(\tau)=Q(\mathcal{T})+\int_0^\tau e^{-2A(\mathcal{T})s}\delta F(\mathcal{T})ds$.
We then obtain the solution:
\begin{equation}
\begin{split}
\delta \Xi^{(0)}&=e^{2A(\mathcal{T})\tau}Q(\mathcal{T})+\int_{0}^{\tau} e^{2(\tau-s)A(\mathcal{T})}\delta F(\mathcal{T}) ds\nonumber\\
&=e^{2A(\mathcal{T})\tau}Q(\mathcal{T})+(e^{2A(\mathcal{T})\tau}-1)\frac{A(\mathcal{T})^{-1}}{2}\delta F(\mathcal{T}).
\end{split}\tag{S3}
\end{equation}
To cancel out the secular term proportional to $\tau$ in $-\frac{\partial \delta \Xi^{(0)}}{\partial \mathcal{T}}$ in the equation of $O(\epsilon)$ in Eq.~(S2), 
we find that we need to choose $Q(\mathcal{T})=-\frac{A(\mathcal{T})^{-1}}{2}\delta F(\mathcal{T})$. Then, we identify
\begin{equation}
\delta \Xi^{(0)}=-\frac{A(\mathcal{T})^{-1}}{2}\delta F(\mathcal{T}).\tag{S4}
\end{equation}
By putting Eq.~(S4) into Eq.~(S2)
and repeating the same procedure
as in the case of $\delta \Xi^{(0)}$, we derive $\delta \Xi^{(1)}$ as
\begin{equation}
\delta \Xi^{(1)}=-\frac{A(\mathcal{T})^{-1}}{2}\frac{\partial}{\partial \mathcal{T}}\left(\frac{A(\mathcal{T})^{-1}}{2}\delta F(\mathcal{T})\right)=\frac{A(\mathcal{T})^{-1}}{2}\frac{\partial \delta \Xi^{(0)}}{\partial \mathcal{T}}.\tag{S5}
\end{equation}
Therefore, we obtain Eqs.~(15)--(17) as
\begin{equation}
\begin{split}
\Xi(t)&=\Xi_0+\delta \Xi(t)\\
&=-\frac{A_0^{-1}}{2}B_0-\frac{A^{-1}}{2}(2\delta A \mathrm \Xi_0+\delta B)+\frac{A^{-1}}{2}\frac{\partial \delta \Xi^{(0)}}{\partial \mathcal T}\epsilon+O(\epsilon^2), \\
&=-\frac{A^{-1}}{2}B+\frac{A^{-1}}{2}\frac{\partial \Xi_{\rm ad}}{\partial \mathcal T}\epsilon+O(\epsilon^2)\\
&\simeq \Xi_{\rm ad}+\delta \Xi_{\rm nad},
\end{split}\tag{S6}
\end{equation}
where we have used $\delta A=A-A_0$, Eq.~(11), and $\frac{\partial \delta \Xi^{(0)}}{\partial \mathcal{T}}=\frac{\partial (\Xi_0+\delta \Xi^{(0)})}{\partial \mathcal{T}}=\frac{\partial \Xi_{\rm ad}}{\partial \mathcal{T}}$ from the second to the third lines.
We note that the adiabatic solution in Eq.~(16) can be obtained by solving Eq.~(10) by formally setting $\partial_t \Xi=0$ from the beginning. Here, we derived it using the time-scale separation method.

\section{Derivation of Eqs.~(22) and (23)}
We can write $\bm j_w=\frac{1}{2}\left<\mathbf x^{\mathrm{T}} \frac{\partial \mathrm H(t)}{\partial \boldsymbol \lambda} \mathbf x\right>=\frac{1}{2}\frac{\partial \mathrm H_{ij}(t)}{\partial \boldsymbol \lambda}\mathrm \Xi_{ij} \simeq \frac{1}{2}\frac{\partial \mathrm H_{ij}(t)}{\partial \boldsymbol \lambda}(\mathrm \Xi_{{\rm ad},ij}+\delta \mathrm \Xi_{{\rm nad},ij})$ as follows: 
\begin{equation}
\bm j_w=-\frac{k_{\mathrm B}T(t)}{2}\mathrm \Xi_{{\rm ad},ij}^{-1}\frac{\partial \mathrm \Xi_{{\rm ad},ij}}{\partial \boldsymbol \lambda}-\frac{k_{\mathrm B}T(t)}{2}\mathrm \Xi_{{\rm ad}, ik}^{-1}\frac{\partial \mathrm \Xi_{{\rm ad}, kl}}{\partial \boldsymbol \lambda}\mathrm \Xi_{{\rm ad}, lj}^{-1}\delta \mathrm \Xi_{{\rm nad},ij},\tag{S7}
\end{equation}
where we have used the relation $\frac{\partial \mathrm H_{ij}(t)}{\partial \boldsymbol \lambda}=-k_{\rm B}T(t)\mathrm \Xi_{{\rm ad},ik}^{-1} \frac{\partial \mathrm \Xi_{{\rm ad},kl}}{\partial \boldsymbol \lambda}\mathrm \Xi_{{\rm ad},lj}^{-1}$ derived from Eq.~(18) and $\mathrm \Xi_{{\rm ad},ik} \mathrm \Xi_{{\rm ad},kj}^{-1}=\delta_{ij}$.
By putting Eq.~(17) into Eq.~(S7) and using $\frac{\partial \mathrm \Xi_{{\rm ad},ij}}{\partial \mathcal{T}}=\frac{\partial \mathrm \Xi_{{\rm ad},ij}}{\partial \boldsymbol \lambda}\cdot \mathbf g'_w(\mathcal{T})$, we obtain Eq.~(22).

We next derive Eq.~(23). 
We can write $j_q=\int d\mathbf x^n \frac{\partial H(\mathbf x, t)}{\partial x_i}\mathcal{J}_i(\mathbf x,t)$ as
\begin{equation}
\begin{split}
j_q&=\int d\mathbf x^n \frac{\partial H(\mathbf x, t)}{\partial x_i}\mathcal{J}_i(\mathbf x,t)\\
&=\int d\mathbf x^n \mathrm{H}_{ij}x_{j}\left(A_{ik}x_k \mathcal{P}(\mathbf x,t)-\frac{1}{2}B_{ik}\frac{\partial \mathcal{P}(\mathbf x,t)}{\partial x_k}\right)\\
&=A_{ik}\mathrm H_{ij}\left<x_jx_k\right>+\frac{1}{2}B_{ij}\mathrm H_{ij}\\
&=A(t)\cdot \mathrm H(t)\Xi(t)+\frac{1}{2}B(t)\cdot \mathrm H(t).\\
\end{split}\tag{S8}
\end{equation}
Then, we have
\begin{equation}
\begin{split}
j_q&=A(t)\cdot \mathrm H(t)\Xi(t)+\frac{1}{2}B(t)\cdot \mathrm H(t)\\
&=k_{\mathrm B}T(t) A(t)\cdot \Xi_{\rm ad}^{-1} (\Xi_{\rm ad}+\delta \Xi_{\rm nad})+\frac{1}{2}B(t)\cdot \mathrm H(t)\\
&=k_{\mathrm B}T(t) A(t)\cdot \left(\mathbf 1-B^{-1}(t)\frac{\partial \Xi_{\rm ad}}{\partial \mathcal{T}}\epsilon\right)+\frac{1}{2}B(t)\cdot \mathrm H(t)\\
&=-k_{\mathrm B}T(t)A_{ij}(t)B_{ik}^{-1}(t)\frac{\partial \Xi_{{\rm ad},kj}}{\partial \mathcal T}\epsilon+k_{\mathrm B}T(t)A_{ii}(t)+\frac{1}{2}B_{ij}(t)H_{ij}(t),
\end{split}\tag{S9}
\end{equation}
where we have used Eq.~(18) from the first to the second lines and Eq.~(17) from the second to the third lines.
By using the detailed balance condition Eq.~(13), $A_{ij}(t)B_{ik}^{-1}(t)=A_{ij}(t)B_{ki}^{-1}(t)=B_{ki}^{-1}(t)A_{ij}(t)=-\frac{1}{2}\Xi_{{\rm ad}, kj}^{-1}$, and $\frac{\partial \mathrm \Xi_{{\rm ad},kj}}{\partial \mathcal{T}}=\frac{\partial \mathrm \Xi_{{\rm ad},kj}}{\partial \boldsymbol \lambda}\cdot \mathbf g'_w(\mathcal{T})$, we obtain Eq.~(23).


\begin{references}
\bibitem{adiabatic_change} The adiabatic change conventionally refers to a thermodynamic process without any heat exchange 
or a sufficiently slow change of parameters compared to relaxation of a system under consideration. In this Letter, we use it as the latter meaning.
\end{references}
\end{document}